\documentclass[pra,twocolumn,showpacs,superscriptaddress]{revtex4-1}
\usepackage{epsfig,graphicx,amssymb,color,bm}
\usepackage{amsmath,float}
\usepackage[colorlinks, linkcolor=blue, urlcolor=blue, citecolor=blue, unicode=true]{hyperref}
\usepackage{setspace}
\usepackage{url}

\begin{document}

\title{The Lindblad and Redfield forms without secular approximation derived from Born-Markov master equation and their applications}
\author{ Chang-Yao Liao, Xian-Ting Liang}
\email[]{liangxianting@nbu.edu.cn}
\affiliation{Department of Physics, Ningbo University, Ningbo 315211, China}
\date{\today}

\begin{abstract}
In this paper we derive the Lindblad and Redfield forms with and without secular approximation from the Born-Markov master equation for open quantum systems. The spectral correlation tensor of bath (the Fourier transform of the bath correlation function) and then the coefficients in the two forms of the master equation are reevaluated according to the scheme in Ref.[Phys. Rev. A 99, 022118 (2019)]. They are complex numbers rather than the real numbers getting from traditional simplified methods. The dynamics of two models [one is an open three-level quantum system model, and the other is the model of  phycoerythrin 545 (PE545) of modeling a photosynthesis reaction center] are studied by using the obtained equations. The non-secular Lindblad and Redfield equations with the complex coefficients predict almost the same dynamical results from the Born-Markov master equation. However, the results obtained from the traditional Lindblad and Redfield equations deviate the actual dynamics of the open quantum systems.
\end{abstract}
\maketitle
\section{Introduction}

Any quantum system will unavoidably suffer from interactions with uncountable degrees of freedom in surroundings. Therefore, the evolution of a quantum state will be affected by the environment of the system. Thus, understanding the dissipative quantum dynamics of a system embedded in a complex environment is an important topic across various sub-disciplines of matter science \cite{UWeiss,RMP59}. The Lindblad and Redfield forms \cite{Redfield1965,Lindblad1976,PRA43} based on Born-Markov master equation are convenient tools and they are widely applied for the investigations of the reduced dynamics of interest quantum systems \cite{JCP96,JCP100,JCP119,PRA75HPB,PRA75SM,PRE78}. However, compared to the Born-Markov master equation, in Lindblad and Redfield forms of the master equation, the secular approximation is always needed to be used. In order to determine the coefficients of the equations, traditionally a Cauchy principle value integral should be solved and it is always difficult to obtain the analytic solution \cite{Breuer2002}. So the imaginary part of the spectral correlation tensor of bath (the Fourier transform of the bath correlation function) is always ignored, which distorts the true dynamics of the open quantum systems. In Ref.\cite{PRA99}, we proposed a scheme of formulating the Born-Markov master equation into a computable form, in which the secular approximation does not need to be used and the coefficients in the master equation are exactly reevaluated. It is interested that based on the idea in Ref.\cite{PRA99}, the Born-markov master equation can be represented into the Lindblad and Redfield forms, in which the secular approximation does not need to be used and the coefficients of the master equations can also be exactly determined. The obtained equations degenerate to the traditional Lindblad and Redfield forms of the master equation when we add the secular approximation on them. For convenience, we call the corresponding equations with and without use of the secular approximation the secular and non-secular Lindblad and Redfield master equations.

By using the non-secular Lindblad and Redfield master equations, in this paper we shall investigate two models. One is a three-level quantum system model, the other is the PE545 complex \cite{nature463,JPCB115,JPCB118} which is a model extensively used to study photosynthetic pigment protein complex. It will be shown that for the three-level quantum system the dynamics obtained from the non-secular Lindblad and Redfield master equations with the complex spectral tensor are similar to ones that obtained from the traditional Lindblad and Redfield master equations. However, for the PE545 complex model, the dynamics obtained from the non-secular forms with the complex spectral correlation tensor are definitely different to the results obtained from the traditional Lindblad and Redfield master equations. It means that for some complex open quantum systems, the traditional Lindblad and Redfield master equations are certainly lose some dynamical information because they omit many relaxation terms which do not satisfy the secular approximation. It also means that the  the imaginary parts of the spectral correlation tensor of bath should also be included in the solution of the master equations.

\section{model and formulation}

In this section we derive the non-secular Lindblad and Redfield equations. As usually, the total system-environment Hamiltonian is set as \cite{Breuer2002,PRA51}

\begin{eqnarray}
  H=H_s+H_b+H_I,
\end{eqnarray}
where $H_s $, $H_b $ and $H_I $ are the Hamiltonian of the system, bath, and their interaction between them. Suppose the coupling of the system and its environment is described by the interaction Hamiltonian as
\begin{eqnarray}
  H_I=\sum_{\alpha}S_{\alpha}\otimes E_{\alpha},\label{HI}
\end{eqnarray}
where $S_{\alpha} $ and $E_{\alpha}$ are the system and environment operators. The general Hamiltonian of the open multi-level quantum system can be written as
\begin{eqnarray}
H=H_s+\sum_k \left[\frac{\hat{p}_k^2}{2m_k} +\frac{1}{2} m_k\omega_k^2\hat{q}_k^2+\sum_\alpha | \alpha\rangle\langle\alpha | c_{k\alpha}\hat{q}_k \right]. \qquad \label{H3}
\end{eqnarray}
Here, $\hat{q}_{k} $, and $\hat{p}_{k} $, are the coordinate and momentum operators, and $m_{k} $ and $\omega_{k}$, the mass and angular frequency of the $k$-th harmonic oscillator of the bath, respectively, and $c_{k \alpha} $ is the coupling coefficient between the $\alpha$-th diagonal mode of the system and the $k$-th harmonic oscillator of the bath. $| \alpha \rangle $ is the $\alpha$-th basis state of the multi-level quantum system. This is actually the Frenkel-exciton Hamiltonian, which is widely used in the study of molecular aggregates in photosynthesis systems and other condensed systems \cite{mukamel1995principles,Science288(2000)1620,JCP134(2011)05B602}. By using the Born-Markov approximation, one can obtain the Born-Markov master equation in Schr$\ddot{o}$dinger picture as  \cite{Breuer2002}
\begin{eqnarray}
\frac{d\rho_s}{dt}&&=-i\left[ H_s,\rho_s\right] +\int_0^{\infty}d\tau \sum_{\alpha\beta}W_{\alpha\beta}(\tau)\nonumber \\
&&\left\lbrace S_{\beta}(-\tau)\rho_s S_{\alpha}-S_{\alpha}S_{\beta}(-\tau)\rho_s \right\rbrace +H.c.. \label{Born-Markov}
\end{eqnarray}
Here, and in the following we set $\hbar=1$, and
\begin{eqnarray}
  &W_{\alpha\beta}(\tau)=Tr_b\left[ E_{\alpha}(\tau)E_{\beta}\rho_b\right], \nonumber \\
  &E_{\alpha}(\tau)=e^{iH_b\tau}E_{\alpha}e^{-iH_b\tau}, \quad S_{\alpha}(\tau)=e^{iH_s\tau}S_{\alpha}e^{-iH_s\tau}. \quad
\end{eqnarray}
Eq.(\ref{Born-Markov}) is not a computable form and it is not convenience for numerical analysis. Fortunately, two computable forms, Lindblad and Redfield forms of the master equation are introduced and extensively used in last years. These equations have been derived in many references, but some of them are based on the secular approximation. In the following we re-derive the equations based on the Born-Markov master equation, and these derivations do not need the help of the secular approximation and other assumptions.
\subsection{Lindblad form}
In this section we derive the Lindblad form of the master equation independent of secular approximation. Setting
\begin{eqnarray}
  L^{0}_{\alpha}=\sum_{\beta}\int_0^{\infty}d\tau S_{\beta}(-\tau)W_{\alpha\beta}(\tau),\label{Lalpha01}
\end{eqnarray}
we can then written Eq.(\ref{Born-Markov}) as
\begin{eqnarray}
\frac{d\rho_s}{dt}&&=-i\left[ H_s,\rho_s\right] +\sum_{\alpha} \left[ L^{0}_{\alpha}\rho_s S_{\alpha}-S_{\alpha}L^{0}_{\alpha}\rho_s\right] \nonumber \\
&&+\left[ S_{\alpha}\rho_s L^{0\dagger}_{\alpha}-\rho_s L^{0\dagger}_{\alpha}S_{\alpha}\right].
\end{eqnarray}
Here, $S_{\alpha} $ is a unitary operator, so $S_{\alpha}=S_{\alpha}^{\dagger}$. Because
\begin{eqnarray}
&&L^{0}_{\alpha}\rho_s S_{\alpha}-S_{\alpha}L^{0}_{\alpha}\rho_s +S_{\alpha}\rho_sL^{0\dagger}_{\alpha}-\rho_s L^{0\dagger}_{\alpha}S_{\alpha} \nonumber \\
&&=\left[ L^{1}_{\alpha}\rho_s L^{1\dagger}_{\alpha}-\frac{1}{2}L^{1\dagger}_{\alpha}L^{1}_{\alpha}\rho_s - \frac{1}{2}\rho_s L^{1\dagger}_{\alpha}L^{1}_{\alpha}\right]  \nonumber \\
&&-\left[ L^{0}_{\alpha}\rho_s L^{0\dagger}_{\alpha}-\frac{1}{2}L^{0\dagger}_{\alpha}L^{0}_{\alpha}\rho_s - \frac{1}{2}\rho_s L^{0\dagger}_{\alpha}L^{0}_{\alpha}\right]  \nonumber \\
&&-\left[ S_{\alpha}\rho_s S^{\dagger}_{\alpha}-\frac{1}{2}S^{\dagger}_{\alpha}S_{\alpha}\rho_s - \frac{1}{2}\rho_s S^{\dagger}_{\alpha}S_{\alpha}\right] \nonumber \\
&&+\frac{1}{2}\left[ L^{0\dagger}_{\alpha}S_{\alpha}-S^{\dagger}_{\alpha}L^{0}_{\alpha},\rho_s\right],
\end{eqnarray}
where
\begin{eqnarray}
  L^{1}_{\alpha}=L^{0}_{\alpha}+S_{\alpha},
\end{eqnarray}
the Born-Markov master equation can be arranged into
\begin{eqnarray}
\frac{d\rho_s}{dt}&&=-i\left[ H_{s}+H_{ls},\rho_s\right]+D(\rho_s), \nonumber \\
\end{eqnarray}
with
\begin{eqnarray}
D(\rho_s)&&=\sum_{\alpha}\left[ L^{1}_{\alpha}\rho_s L^{1\dagger}_{\alpha}-\frac{1}{2}L^{1\dagger}_{\alpha}L^{1}_{\alpha}\rho_s - \frac{1}{2}\rho_s L^{1\dagger}_{\alpha}L^{1}_{\alpha}\right]  \nonumber \\
&&-\left[ L^{0}_{\alpha}\rho_s L^{0\dagger}_{\alpha}-\frac{1}{2}L^{0\dagger}_{\alpha}L^{0}_{\alpha}\rho_s - \frac{1}{2}\rho_s L^{0\dagger}_{\alpha}L^{0}_{\alpha}\right]  \nonumber \\
&&-\left[ S_{\alpha}\rho_s S^{\dagger}_{\alpha}-\frac{1}{2}S^{\dagger}_{\alpha}S_{\alpha}\rho_s - \frac{1}{2}\rho_s S^{\dagger}_{\alpha}S_{\alpha}\right],\label{Drhos}
\end{eqnarray}
and
\begin{eqnarray}
  H_{ls}=\frac{i}{2}\sum_{\alpha}\left[ L^{0\dagger}_{\alpha}S_{\alpha}-S^{\dagger}_{\alpha}L^{0}_{\alpha}\right].
\end{eqnarray}
We denote the eigenvalues of $H_s $ by $\omega_n$ and the projection onto the eigenspace belonging to the eigenvalue $\omega_n $ by $\Pi (\omega_n) $. Then, we define the operators \citep{Breuer2002}
\begin{eqnarray}
  F_{\alpha}(\omega_{mn})\equiv\Pi (\omega_n)S_{\alpha}\Pi (\omega_m). \label{Falpha}
\end{eqnarray}
Summing over all energy levels and employing the completeness relation in Eq.(\ref{Falpha}), we get
\begin{eqnarray}
  S_{\alpha}=\sum_{mn}F_{\alpha}(\omega_{mn}).\label{Salpha1}
\end{eqnarray}
Thus, we have
\begin{eqnarray}
S_{\alpha}(\tau)&&=e^{iH_s\tau}S_{\alpha}e^{-iH_s\tau}\nonumber\\
&&=\sum_{mn}e^{iH_s\tau}F_{\alpha}(\omega_{mn})e^{-iH_s\tau}\nonumber\\
&&=\sum_{mn}e^{-i\omega_{mn}\tau}F_{\alpha}(\omega_{mn}). \quad \label{Salpha2}
\end{eqnarray}
Inserting the form of Eq.(\ref{Salpha2}) into the Eq.(\ref{Lalpha01}), we obtain
\begin{eqnarray}
L^{0}_{\alpha}&&=\sum_{\beta}\int_0^{\infty}d\tau S_{\beta}(-\tau)W_{\alpha\beta}(\tau)\nonumber\\
&&=\sum_{\beta}\int_0^{\infty}d\tau \sum_{{mn}}e^{i\omega_{mn}\tau}F_{\beta}(\omega_{mn})W_{\alpha\beta}(\tau)\nonumber\\
&&=\sum_{\beta,mn}\Gamma_{\alpha\beta}(\omega_{mn})F_{\beta}(\omega_{mn}),\label{Lalpha02}
\end{eqnarray}
where, $\Gamma_{\alpha\beta}(\omega_{mn}) $ is the Fourier transform of the bath correlation function, we call it the spectral correlation tensor of bath, and it is
\begin{eqnarray}
  \Gamma_{\alpha\beta}(\omega_{mn})\equiv\int_0^{\infty}dse^{i\omega_{mn} s}W_{\alpha\beta}(s).
\end{eqnarray}
So, we have
\begin{eqnarray}
  L^1_{\alpha}=L^0_{\alpha}+S_{\alpha}=\sum_{\beta,mn}\left( \Gamma_{\alpha\beta}(\omega_{mn})+\delta_{\alpha\beta}\right) F_{\beta}(\omega_{mn}).\label{Lalpha1}\quad
\end{eqnarray}
Inserting Eqs.(\ref{Salpha1}), (\ref{Lalpha02}), and (\ref{Lalpha1}) into Eq.(\ref{Drhos}) we can obtain,
\begin{eqnarray}
&&L_{\alpha}^1\rho_s L_{\alpha}^{1\dagger}-\frac{1}{2}L_{\alpha}^{1\dagger}L_{\alpha}^1\rho_s-\frac{1}{2}\rho_sL_{\alpha}^{1\dagger}L_{\alpha}^1 \nonumber \\
&&=\sum_{s^{\prime}s}\sum_{mn}\sum_{\beta\beta^{\prime}}\left[ \left( \delta_{\alpha\beta}+\Gamma_{\alpha\beta}(\omega_{s^{\prime}s})\right) \left( \delta_{\alpha\beta^{\prime}}+\Gamma^*_{\beta^{\prime}\alpha}(\omega_{mn})\right)\right]\nonumber \\
&&[ F_{\beta}(\omega_{s^{\prime}s})\rho_s F^{\dagger}_{\beta^{\prime}}(\omega_{mn})-\frac{1}{2}F^{\dagger}_{\beta^{\prime}}(\omega_{mn})F_{\beta}(\omega_{s^{\prime}s})\rho_s \nonumber \\
&&-\frac{1}{2}\rho_s F^{\dagger}_{\beta^{\prime}}(\omega_{mn})F_{\beta}(\omega_{s^{\prime}s})],
\end{eqnarray}
and
\begin{eqnarray}
&&L_{\alpha}^0\rho_s L_{\alpha}^{0\dagger}-\frac{1}{2}L_{\alpha}^{0\dagger}L_{\alpha}^0\rho_s-\frac{1}{2}\rho_sL_{\alpha}^{0\dagger}L_{\alpha}^0 \nonumber \\
&&=\sum_{s^{\prime}s}\sum_{mn}\sum_{\beta\beta^{\prime}}\left[ \Gamma_{\alpha\beta}(\omega_{s^{\prime}s})\Gamma^*_{\beta^{\prime}\alpha}(\omega_{mn})\right]\nonumber \\
&&\times [ F_{\beta}(\omega_{s^{\prime}s})\rho_s F^{\dagger}_{\beta^{\prime}}(\omega_{mn})-\frac{1}{2}F^{\dagger}_{\beta^{\prime}}(\omega_{mn})F_{\beta}(\omega_{s^{\prime}s})\rho_s \nonumber \\
&&-\frac{1}{2}\rho_s F^{\dagger}_{\beta^{\prime}}(\omega_{mn})F_{\beta}(\omega_{s^{\prime}s})],
\end{eqnarray}
and
\begin{eqnarray}
&&S_{\alpha}\rho_s S_{\alpha}^{\dagger}-\frac{1}{2}S_{\alpha}^{\dagger}S_{\alpha}\rho_s -\frac{1}{2}\rho_s S_{\alpha}^{\dagger}S_{\alpha} \nonumber \\
&&=\sum_{s^{\prime}s}\sum_{mn}\sum_{\beta\beta^{\prime}}[ F_{\beta}(\omega_{s^{\prime}s})\rho_s F^{\dagger}_{\beta^{\prime}}(\omega_{mn})\nonumber\\
&&-\frac{1}{2}F^{\dagger}_{\beta^{\prime}}(\omega_{mn})F_{\beta}(\omega_{s^{\prime}s})\rho_s-\frac{1}{2}\rho_s F^{\dagger}_{\beta^{\prime}}(\omega_{mn})F_{\beta}(\omega_{s^{\prime}s})]. \quad
\end{eqnarray}
Thus, we have
\begin{eqnarray}
&&D(\rho_s)=\sum_{s^{\prime}s}\sum_{mn}\sum_{\beta\beta^{\prime}}\left( \Gamma^*_{\beta^{\prime}\beta}(\omega_{mn})+\Gamma_{\beta^{\prime}\beta}(\omega_{s^{\prime}s})\right) \nonumber \\
&&[ F_{\beta}(\omega_{s^{\prime}s})\rho_s F^{\dagger}_{\beta^{\prime}}(\omega_{mn})-\frac{1}{2}F^{\dagger}_{\beta^{\prime}}(\omega_{mn})F_{\beta}(\omega_{s^{\prime}s})\rho_s \nonumber \\
&&-\frac{1}{2}\rho_s F^{\dagger}_{\beta^{\prime}}(\omega_{s^{\prime}s})F_{\beta}(\omega_{mn})].\label{Drhos2}
\end{eqnarray}
Similarly, we have
\begin{eqnarray}
H_{ls}&&=\frac{i}{2}\sum_{\alpha}\left[ L^{0\dagger}_{\alpha}S_{\alpha}-S^{\dagger}_{\alpha}L^{0}_{\alpha}\right] \nonumber \\
&&=\frac{i}{2}\sum_{\alpha\beta}\sum_{s^{\prime}s}\sum_{mn}[ \Gamma^*_{\beta\alpha}(\omega_{s^{\prime}s})F^{\dagger}_{\beta}(\omega_{s^{\prime}s})F_{\alpha}(\omega_{mn}) \nonumber \\
&&-\Gamma_{\alpha\beta}(\omega_{s^{\prime}s})F^{\dagger}_{\alpha}(\omega_{mn})F_{\beta}(\omega_{s^{\prime}s})] \nonumber \\
&&=\frac{i}{2}\sum_{\alpha\beta}\sum_{s^{\prime}s}\sum_{mn}[ \Gamma^*_{\alpha\beta}(\omega_{mn})F^{\dagger}_{\alpha}(\omega_{mn})F_{\beta}(\omega_{s^{\prime}s}) \nonumber \\
&&-\Gamma_{\alpha\beta}(\omega_{s^{\prime}s})F^{\dagger}_{\alpha}(\omega_{mn})F_{\beta}(\omega_{s^{\prime}s})] \nonumber \\
&&=\frac{i}{2}\sum_{\alpha\beta}\sum_{s^{\prime}s}\sum_{mn}\left[ \Gamma^*_{\alpha\beta}(\omega_{mn})-\Gamma_{\alpha\beta}(\omega_{s^{\prime}s})\right]\nonumber \\
&&\times F^{\dagger}_{\alpha}(\omega_{mn})F_{\beta}(\omega_{s^{\prime}s}).  \label{Hls}
\end{eqnarray}
In Eq.(\ref{Hls}), the third equality holds, which is based on the fact that exchanging the indexes $\alpha$ and $\beta $, and variables $\omega_{s^{\prime}s}$ and $\omega_{mn}$, the first term of the right-hand side of the equality does not change. We denote,
\begin{eqnarray}
  \Gamma_{\alpha\beta}(\omega_{mn})= \frac{1}{2}\gamma_{\alpha\beta}(\omega_{mn})+i T_{\alpha\beta}(\omega_{mn}).
\end{eqnarray}
So, we have
\begin{eqnarray}
&&\Gamma^*_{\alpha \beta}(\omega_{mn})+\Gamma_{\alpha \beta}(\omega_{s^{\prime}s}) \nonumber \\
&&=\frac{1}{2}\left[\gamma_{\alpha\beta}(\omega_{mn})+\gamma_{\alpha\beta}(\omega_{s^{\prime}s}))-i (T_{\alpha\beta}(\omega_{mn})- T_{\alpha\beta}(\omega_{s^{\prime}s})\right] \nonumber \\
&&\equiv\chi_{\alpha\beta}(\omega_{mn},\omega_{s^{\prime}s}), \label{11}
\end{eqnarray}
and
\begin{eqnarray}
&&\Gamma^*_{\alpha\beta}(\omega_{s^{\prime}s})-\Gamma_{\alpha\beta}(\omega_{mn}) \nonumber \\
&&=\frac{1}{2}\left[\gamma_{\alpha\beta}(\omega_{mn})-\gamma_{\alpha\beta}(\omega_{s^{\prime}s}))-i (T_{\alpha\beta}(\omega_{mn})+ T_{\alpha\beta}(\omega_{s^{\prime}s})\right] \nonumber \\
&&\equiv\Theta_{\alpha\beta}(\omega_{mn},\omega_{s^{\prime}s}). \label{22}
\end{eqnarray}
Thus, the master equation can be written in the compact form as
\begin{eqnarray}
  \frac{d}{dt}\rho_s(t)=-i\left[ H_s+H_{ls}\right] +D(\rho_s), \label{Lindblad}
\end{eqnarray}
with
\begin{eqnarray}
  H_{ls}=\frac{i}{2}\sum_{\alpha\beta}\sum_{s^{\prime}s}\sum_{mn} \Theta_{\alpha\beta}(\omega_{mn},\omega_{s^{\prime}s})F^{\dagger}_{\alpha}(\omega_{mn})F_{\beta}(\omega_{s^{\prime}s}),\qquad
\end{eqnarray}
and
\begin{eqnarray}
&&D(\rho_s)=\sum_{\alpha\beta}\chi_{\alpha\beta}(\omega_{mn},\omega_{s^{\prime}s})[ F_{\beta}(\omega_{s^{\prime}s})\rho_s F^{\dagger}_{\alpha}(\omega_{mn}) \nonumber \\
  &&-\frac{1}{2}F^{\dagger}_{\alpha}(\omega_{mn})F_{\beta}(\omega_{s^{\prime}s})\rho_s-\frac{1}{2}\rho_s F^{\dagger}_{\alpha}(\omega_{mn})F_{\beta}(\omega_{s^{\prime}s})]. \label{D}
\end{eqnarray}
Further, this formation degenerates into the traditional form of the Lindblad master equation through imposing the secular approximation, and it becomes into \cite{JCP129,JMP17}
\begin{eqnarray}
  \frac{d}{dt}\rho_s(t)=-i\left[ H_s+H^{se}_{ls}\right] +D^{se}(\rho_s),\label{Lindbladse}
\end{eqnarray}
with
\begin{eqnarray}
  H^{se}_{ls}=\frac{i}{2}\sum_{\alpha\beta}\sum_{\omega_{mn}}\Theta_{\alpha\beta}(\omega_{mn},\omega_{mn})F^{\dagger}_{\alpha}(\omega_{mn})F_{\beta}(\omega_{mn}), \nonumber \\
\end{eqnarray}
and
\begin{eqnarray}
 &&D^{se}(\rho)=\sum_{\omega_{mn}}\sum_{\alpha\beta}\chi_{\alpha\beta}(\omega_{mn},\omega_{mn})[ F_{\beta}(\omega_{mn})\rho_s F^{\dagger}_{\alpha}(\omega_{mn}) \nonumber \\
 &&-\frac{1}{2}F^{\dagger}_{\alpha}(\omega_{mn})F_{\beta}(\omega_{mn})\rho_s-\frac{1}{2}\rho_s F^{\dagger}_{\alpha}(\omega_{mn})F_{\beta}(\omega_{mn})].\label{Dse}
\end{eqnarray}
Thus, we obtain the Lindblad form of the master equation with and without secular approximation.

\subsection{Redfiled form}

Similary, in the following we drive the non-secular Redfield equation. It is known that in the inteaction picture, we have
\begin{eqnarray}
  &&\rho_s^I(t)=e^{iH_st}\rho_s e^{-iH_st}, \quad H_I^I(t)=\sum_{\alpha}S^I_{\alpha}(t)\otimes E^I_{\alpha}(t), \nonumber \\
  &&S^I_{\alpha}(t)=e^{iH_st}S_{\alpha}e^{-iH_st}, \quad E^I_{\alpha}(t)=e^{iH_bt}E_{\alpha}e^{-iH_bt}. \label{inter}
\end{eqnarray}
From Eqs.(\ref{Born-Markov}) and (\ref{inter}) we have
\begin{eqnarray}
 \frac{\partial}{\partial}\rho_s^I(t)=&&-\sum_{\alpha\beta}\int_0^td\tau\left [S_{\alpha}^I(t),S_{\beta}^I(t-\tau)\rho_s^I(t) \right ]W_{\alpha\beta}(\tau) \nonumber \\
  &&-\left [S_{\alpha}^I(t),\rho_s^I(t)S_{\beta}^I(t-\tau) \right ]W_{\beta\alpha}(-\tau). \label{B1}
\end{eqnarray}
The matrix form of Eq.(\ref{B1}) reads
\begin{eqnarray}
  \langle s^{\prime}|\frac{\partial\rho_s^I(t)}{\partial t}|s\rangle =&& -\sum_{\alpha\beta}\int_0^{\infty}d\tau M_{s^{\prime}s}^{(1)}(\alpha ,\beta ,t ,\tau)W_{\alpha\beta}(\tau) \nonumber \\
  &&-M_{s^{\prime}s}^{(2)}(\alpha ,\beta ,t ,\tau)W_{\beta\alpha}(-\tau),
\end{eqnarray}
with
\begin{eqnarray}
  &&M_{s^{\prime}s}^{(1)}(\alpha ,\beta ,t ,\tau)=\langle s^{\prime}|\left [S_{\alpha}^I(t),S_{\beta}^I(t-\tau)\rho_s^I(t) \right ] |s\rangle \nonumber \\
  &&=\sum_{mn}\langle s^{\prime}|S_{\alpha}^I(t)|m\rangle\langle m|S_{\beta}^I(t-\tau)|n\rangle\langle n|\rho_s^I(t)|s\rangle \nonumber \\
  &&-\sum_{mn}\langle s^{\prime}|S_{\beta}^I(t-\tau)|m\rangle\langle m|\rho_s^I(t)|n\rangle\langle n|S_{\alpha}^I(t)|s\rangle \nonumber \\
  &&=\sum_{mn}\langle s^{\prime}|S_{\alpha}|m\rangle\langle m|S_{\beta}|n\rangle\langle n|\rho_s^I(t)|s\rangle e^{i\omega_{s^{\prime}n}t}e^{-i\omega_{mn}\tau}  \nonumber \\
  &&-\sum_{mn}\langle n|S_{\alpha}|s\rangle\langle s^{\prime}|S_{\beta}|m\rangle\langle m|\rho_s^I(t)|n\rangle e^{i(\omega_{s^{\prime}s}+\omega_{nm})t}e^{-i\omega_{s^{\prime}m}\tau}  \nonumber \\
  &&=\sum_{mn}m^{\alpha \beta}_{s^{\prime}mmn} \langle n|\rho_s^I(t)|s\rangle e^{i\omega_{s^{\prime}n}t}e^{-i\omega_{mn}\tau}\nonumber \\
  &&-m^{\alpha\beta}_{nss^{\prime}m}\langle m|\rho_s^I(t)|n\rangle e^{i(\omega_{s^{\prime}s}+\omega_{nm})t}e^{-i\omega_{s^{\prime}m}\tau},
\end{eqnarray}
and
\begin{eqnarray}
  &&M_{s^{\prime}s}^{(2)}(\alpha ,\beta ,t ,\tau)=\langle s^{\prime}|\left [S_{\alpha}^I(t),\rho_s^I(t)S_{\beta}^I(t-\tau) \right ] |s\rangle \nonumber \\
  &&=\sum_{mn}m^{\alpha \beta}_{s^{\prime}mns} \langle m|\rho_s^I(t)|n\rangle e^{i(\omega_{s^{\prime}s}+\omega_{nm})t}e^{-i\omega_{ns}\tau} \nonumber \\
  &&-m^{\alpha\beta}_{nsmn}\langle s^{\prime}|\rho_s^I(t)|m\rangle e^{i\omega_{ms}t}e^{-i\omega_{mn}\tau},
\end{eqnarray}
where
\begin{eqnarray}
  &&m^{\alpha \beta}_{s^{\prime}smn}=\langle s^{\prime}|S_{\alpha}|s\rangle\langle m|S_{\beta}|n\rangle \nonumber \\
  &&\omega_{s^{\prime}s}=\omega_{s^{\prime}}-\omega_s .
\end{eqnarray}
Thus, we have
\begin{eqnarray}
  \langle s^{\prime}|\frac{\partial\rho_s^I(t)}{\partial t}|s\rangle &&=-\sum_{mn}\left[ \Gamma_{s^{\prime}mmn}^+\langle n|\rho_s^I(t)|s\rangle e^{i\omega_{s^{\prime}n}t}\right. \nonumber \\
  &&-\Gamma_{nss^{\prime}m}^+\langle m|\rho_s^I(t)|n\rangle e^{i(\omega_{s^{\prime}m}+\omega_{ns})t} \nonumber \\
  &&-\Gamma_{s^{\prime}mns}^-\langle m|\rho_s^I(t)|n\rangle e^{i(\omega_{s^{\prime}s}+\omega_{nm})t} \nonumber \\
  && \left.+\Gamma_{nsmn}^-\langle s^{\prime}|\rho_s^I(t)|m\rangle e^{i\omega_{ms}t}\right] ,  \label{Lindblad1}
\end{eqnarray}
where
\begin{eqnarray}
  \Gamma_{s^{\prime}mmn}^+&&=\sum_{\alpha\beta}\int_0^td\tau m^{\alpha\beta}_{s^{\prime}mmn}W_{\alpha\beta}(\tau)e^{i\omega_{nm}\tau} \nonumber \\
  &&=\sum_{\alpha\beta}m^{\alpha\beta}_{s^{\prime}mmn}\Gamma_{\alpha\beta}(\omega_{nm}), \nonumber \\
  \Gamma_{nss^{\prime}m}^+&&=\sum_{\alpha\beta}\int_0^td\tau m^{\alpha\beta}_{nss^{\prime}m}W_{\alpha\beta}(\tau)e^{i\omega_{ms^{\prime}}\tau} \nonumber \\
  &&=\sum_{\alpha\beta}m^{\alpha\beta}_{nss^{\prime}m}\Gamma_{\alpha\beta}(\omega_{ms^{\prime}}), \nonumber \\
  \Gamma_{s^{\prime}mns}^-&&=\sum_{\alpha\beta}\int_0^td\tau m^{\alpha\beta}_{s^{\prime}mns}W_{\alpha\beta}(-\tau)e^{-i\omega_{ns}\tau} \nonumber \\
  &&=\sum_{\alpha\beta}m^{\alpha\beta}_{s^{\prime}mns}\Gamma_{\alpha\beta}^*(\omega_{ns}), \nonumber \\
  \Gamma_{nsmn}^-&&=\sum_{\alpha\beta}\int_0^td\tau m^{\alpha\beta}_{nsmn}W_{\alpha\beta}(-\tau)e^{-i\omega_{mn}\tau} \nonumber \\
  &&=\sum_{\alpha\beta}m^{\alpha\beta}_{nsmn}\Gamma_{\alpha\beta}^*(\omega_{mn}).  \label{GammapGammam1}
\end{eqnarray}
By using Eq.(\ref{GammapGammam1}) we have
\begin{eqnarray}
  &&\sum_{mn}\Gamma_{s^{\prime}mmn}^+\langle n|\rho_s^I(t)|s\rangle e^{i\omega_{s^{\prime}n}t} \nonumber \\
  &&=\sum_{nk}\Gamma_{s^{\prime}kkn}^+\langle n|\rho_s^I(t)|s\rangle e^{i\omega_{s^{\prime}n}t} \nonumber \\
  &&=\sum_{mk}\Gamma_{s^{\prime}kkm}^+\langle m|\rho_s^I(t)|s\rangle e^{i(\omega_{s^{\prime}s}-\omega_{ms})t} \nonumber \\
  &&=\sum_{mnk}\Gamma_{s^{\prime}kkm}^+\delta_{sn}\langle m|\rho_s^I(t)|n\rangle e^{i(\omega_{s^{\prime}s}-\omega_{mn})t}, \label{Gammap2}
\end{eqnarray}
and
\begin{eqnarray}
  &&\sum_{mn}\Gamma_{nsmn}^-\langle s^{\prime}|\rho_s^I(t)|m\rangle e^{i\omega_{ms}t} \nonumber \\
  &&=\sum_{mk}\Gamma_{ksmk}^-\langle s^{\prime}|\rho_s^I(t)|m\rangle e^{i\omega_{ms}t} \nonumber \\
  &&=\sum_{nk}\Gamma_{ksnk}^-\langle s^{\prime}|\rho_s^I(t)|n\rangle e^{i(\omega_{ms}-\omega_{mn})t} \nonumber \\
  &&=\sum_{mnk}\Gamma_{ksnk}^-\delta_{s^{\prime}m}\langle m|\rho_s^I(t)|n\rangle e^{i(\omega_{s^{\prime}s}-\omega_{mn})t}.\label{Gammam2}
\end{eqnarray}
Substituting Eqs.(\ref{Gammap2}), and  (\ref{Gammam2}), into Eq.(\ref{Lindblad1}), we can finally obtain the Redfield equation in the interaction picture as
\begin{eqnarray}
  \langle s^{\prime}|\frac{\partial\rho_s^I(t)}{\partial t}|s\rangle =-\sum_{mn}R_{s^{\prime}smn}^I\langle m| \rho_s^I(t)|n\rangle,
\end{eqnarray}
with
\begin{eqnarray}
  R_{s^{\prime}smn}^I&&=\left[\sum_k\delta_{sn}\Gamma^+_{s^{\prime}kkm}-\Gamma_{nss^{\prime}m}^+\right. \nonumber \\
  &&\left.-\Gamma_{s^{\prime}mns}^-+\sum_k\delta_{s^{\prime}m}\Gamma_{ksnk}^-\right] e^{i(\omega_{s^{\prime}s}-\omega_{mn})t}.
\end{eqnarray}
Thus, coming back to the Schr$\ddot{o} $dinger picture, we can obtain the Redfield equation as
\begin{eqnarray}
  \langle s^{\prime}|\frac{\partial\rho_s}{\partial t}|s\rangle =-i\omega_{s^{\prime}s}\langle s^{\prime}|\rho_s|s\rangle-\sum_{mn}R_{s^{\prime}smn}\langle m| \rho_s|n\rangle,\quad \label{Redfield}
\end{eqnarray}
with
\begin{eqnarray}
  R_{s^{\prime}smn}=\sum_k\delta_{sn}\Gamma^+_{s^{\prime}kkm}-\Gamma_{nss^{\prime}m}^+
  -\Gamma_{s^{\prime}mns}^{-}+\sum_k\delta_{s^{\prime}m}\Gamma_{ksnk}^{-}.\nonumber \\
\label{R}
\end{eqnarray}
This is in fact the non-secular Redfield equation. The form is exactly equaled to the one obtained from the secular approximation. If we set the sum in Eq.(\ref{Redfield}) only including these terms, $\omega_s=\omega_{s^{\prime}}, \omega_m=\omega_{n} $, and $\omega_s-\omega_{s^{\prime}}=\omega_m-\omega_{n}$, the equation then degenerates to secular Redfield equation.

\section{On the coefficients in Lindblad and Redfield master equations}
We can see that the Lindblad and Redfield equations Eqs.(\ref{Lindblad}), (\ref{Lindbladse}) and (\ref{Redfield}) are depended on the coefficients $\chi_{\alpha\beta}(\omega_{mn},\omega_{s^{\prime}s}), \Theta_{\alpha\beta}(\omega_{mn},\omega_{s^{\prime}s})$ and $\Gamma^{\pm}_{ijkl}$ which are depended on ascertain $\Gamma_{\alpha\beta}(\Delta) $ for an identified $\Delta$. In the following, we reevaluate the quality according to the scheme of calculating coefficients of Born-markov master equation in Ref.\cite{PRA99}. It is known that when $\alpha\neq\beta $, we have $W_{\alpha\beta}(\tau)=0$, and when $\alpha=\beta $, we set $W_{\alpha}(\tau)=W_{\alpha\beta}(\tau)$. So the correlate function can be calculated as
\begin{eqnarray}
  &&W_{\alpha}(\tau)=\sum_{j,k}c^*_{j\alpha}c_{k\alpha}\langle q_{j}(\tau)q_{k}\rangle_{\rho_b}=\sum_{j}|c_{j\alpha}|^2\langle q_{j}(\tau)q_{j}\rangle_{\rho_b} \nonumber  \\
  &&=\sum_{j}\frac{|c_{j\alpha}|^2}{2m_{j}\omega_{j}}\left [(1+N(\omega_{j}))e^{-i\omega_{j}\tau}+N(\omega_{j})e^{i\omega_{j}\tau} \right ], \quad
\end{eqnarray}
where we noticed that $\langle q_j(\tau)|q_k\rangle=0$ for $j\ne k$, and $N(\omega_{j})=1/\left[ e^{\omega_{j}/k_BT}-1\right]  $, with $k_B$ the Boltzmann constant, and $T $ the temperature. Traditionally, from $W_{\alpha\beta}(\tau)$ we can obtain the spectral correlation tensor of bath for an identified $\Delta$ as
\begin{eqnarray}
  \Gamma^{(1)}_{\alpha\beta}(\Delta)&&=\int_0^{\infty}d\tau e^{i\Delta\tau}W_{\alpha\beta}(\tau)=0\qquad (\alpha\neq\beta),  \nonumber \\
  \Gamma^{(1)}_{\alpha\beta}(\Delta)&&=\int_0^{\infty}d\tau e^{i\Delta\tau}W_{\alpha}(\tau)  \qquad (\alpha=\beta)\nonumber \\
  &&=\sum_{j}\int_0^{\infty}d\tau\frac{|c_{\alpha j}|^2}{2m_{j}\omega_{j}}  \nonumber \\
  &&\left [(1+N(\omega_{j}))e^{i(\Delta-\omega_{j})\tau}+N(\omega_{j})e^{i(\Delta+\omega_{j})\tau} \right ] \nonumber \\
  &&=\int_0^{\infty}d\tau\int_0^{\infty}d\omega J_{\alpha}(\omega)  \nonumber \\
  &&\left [(1+N(\omega))e^{i(\Delta-\omega)\tau}+N(\omega)e^{i(\Delta+\omega)\tau} \right ] \nonumber \\
  &&\approx \int_0^{\infty}d\omega \pi J_{\alpha}(\omega) \nonumber \\
  &&\left [(1+N(\omega))\delta(\Delta-\omega)+N(\omega)\delta(\Delta+\omega) \right ].\label{Gamma1}
\end{eqnarray}
Here
\begin{eqnarray}
J_{\alpha}(\omega)=\sum_{j}\frac{|c_{\alpha j}|^2}{2m_{j}\omega_{j}}\delta(\omega_{j}-\omega).
\end{eqnarray}
It is always described with a environmentally spectral density function, for example, Drude spectral density function \cite{PRA43}, as
\begin{eqnarray}
J_{\alpha}(\omega)=J(\omega)=\frac{\eta\Omega\omega}{\omega^2+\Omega^2}. \label{J}
\end{eqnarray}
Suppose the baths to be coupled to all modes of the system are the same, and ignore the imaginary part of integral in Eq.(\ref{Gamma1}), an principle-value integral, then $\Gamma_{\alpha\beta}(\omega_{mn})$ in Eqs.(\ref{Drhos2}), (\ref{Hls}) and (\ref{GammapGammam1}) can be obtained as
\begin{eqnarray}
\Gamma_{\alpha\beta}(\omega_{mn})&=&\Gamma_{\alpha\beta}^{(1)}(\omega_{mn})
\end{eqnarray}
Thus, when $\alpha\neq\beta $, $\Gamma^{(1)}_{\alpha\beta}(\omega_{mn})=0 $, and when $\alpha=\beta $ we have
\begin{eqnarray}
  \Gamma^{(1)}_{\alpha\alpha}(\omega_{mn}) = \left\{\begin{array}{ll}
      \pi J(\omega_{mn})(1+N(\omega_{mn})),  &  \textrm{if $\omega_{mn} >0 $,} \\
      \pi J(-\omega_{mn})N(-\omega_{mn}),  &  \textrm{if $\omega_{mn} <0 $,} \\
      \pi \mathop{lim}\limits_{\omega_{mn}\to 0}J(-\omega_{mn})N(-\omega_{mn}), &  \textrm{if $\omega_{mn} =0 $.}
  \end{array} \right.\qquad
\end{eqnarray}
Traditionally, we calculate the coefficients $\chi_{\alpha\beta}(\omega_{mn},\omega_{s^{\prime}s})$, $ \Theta_{\alpha\beta}(\omega_{mn},\omega_{s^{\prime}s})$ and $\Gamma^{\pm}_{ijkl}$ by using the $\Gamma_{\alpha\alpha}^{(1)}(\omega_{mn})$.

However, according to Ref.\cite{PRA99}, we can recalculate the spectral correlation tensor  of bath, and this calculation does not need to ignore the imaginary part, here we set
\begin{eqnarray}
W_{\alpha}(\tau)=\sum_{j}\vert c_{j\alpha}\vert ^2\langle q_{j}(\tau)q_{j}\rangle_{\rho_b}\equiv\nu_{\alpha}(\tau)-i\mu_{\alpha}(\tau), \nonumber \\
\end{eqnarray}
with
\begin{eqnarray}
  \nu_{\alpha}(\tau)&&=\frac{1}{2}\sum_{j}|c_{j\alpha}|^2\langle \left\lbrace q_{j}(\tau)q_{j}\right\rbrace \rangle_{\rho_b} \nonumber \\
  &&=\int_0^{\infty}d\omega J(\omega)\coth(\frac{\omega}{2k_BT})\cos(\omega\tau)=\nu(\tau), \nonumber \\
  \mu_{\alpha}(\tau)&&=\frac{i}{2}\sum_{j}|c_{\alpha j}|^2\langle \left[ q_{j}(\tau)q_{j}\right] \rangle_{\rho_b} \nonumber \\
  &&=\int_0^{\infty}d\omega J(\omega)\sin(\omega\tau)=\mu(\tau).
\end{eqnarray}
Thus, we can recalculate the $\Gamma_{\alpha\beta}(\omega_{mn}) $. When $\alpha\neq\beta $, we have $\Gamma_{\alpha\beta}(\omega_{mn})=0$, and when $\alpha=\beta$, we have
\begin{eqnarray}
&&\Gamma_{\alpha\alpha}(\omega_{mn})=\Gamma_{\alpha\alpha}^{(2)}(\omega_{mn})\nonumber\\
&&=\int_0^{\infty}d\tau e^{i\omega_{mn}\tau}\left[ \nu(\tau)-i\mu(\tau)\right] \nonumber \\
&&=\int_0^{\infty}d\tau\left (\cos(\omega_{mn}\tau)+i\sin(\omega_{mn}\tau)\right )\left (\nu(\tau)-i\mu(\tau) \right ) \nonumber \\
  &&=\bar{D}(\omega_{mn})+i\bar{f}(\omega_{mn})-i\bar{\kappa}(\omega_{mn})+\bar{r}(\omega_{mn}),
\end{eqnarray}
with
\begin{eqnarray}
\bar{D}(\omega_{mn})&&=\int_0^{\infty}d\tau\nu(\tau)\cos(\omega_{mn}\tau) \nonumber \\
&&= \left\{\begin{array}{ll}
      \frac{\pi}{2}J(\omega_{mn})\coth\left( \frac{\omega_{mn}}{2k_BT}\right),  \quad  \textrm{for $\omega_{mn} \neq 0 $}, \\
      \frac{\pi\eta}{2}\cot(\frac{\Omega}{2k_BT}) \\
      +\sum\limits_j^N\frac{2\pi\eta\Omega k_BT}{(\bar{\omega}_j^2-\Omega^2)}, \quad  \textrm{for $\omega_{mn} = 0, \Omega\neq 2k\pi k_BT $}, \\
      -\frac{k_BT\eta\pi}{2\Omega} \\
      +\sum \limits_{j\neq k}^N\frac{2\pi\eta\Omega k_BT}{(\bar{\omega}_j^2-\Omega^2)}, \quad \textrm{for $\omega_{mn} = 0, \Omega= 2k\pi k_BT $}.
  \end{array} \right.\nonumber\\
\bar{f}(\omega_{mn})&&=\int_0^{\infty}d\tau\nu(\tau)\sin(\omega_{mn}\tau) \nonumber \\
&&= \left\{\begin{array}{ll}
    \frac{\pi\eta\Omega\omega_{mn}}{2(\Omega^2+\omega_{mn}^2)}\cot\left(\frac{\Omega}{2 k_B T}\right) \\
    +\sum\limits_j^N \frac{\eta \Omega \bar{\omega}^2_j  \omega_{mn}}{j(\bar{\omega}_j^2-\Omega^2)(\bar{\omega}_j^2+\omega_{mn}^2)}, &\textrm{for $ \Omega\neq 2k\pi k_BT $}, \\
    \frac{\eta\pi}{2\beta}\frac{\omega_{mn}^3-3\omega_{mn}\Omega^2}{\left(\omega_{mn}^2+\Omega^2\right)^2} \\
    +\sum\limits_{j\neq k}^N \frac{\eta \Omega \bar{\omega}^2_j  \omega_{mn}}{j(\bar{\omega}_j^2-\Omega^2)(\bar{\omega}_j^2+\omega_{mn}^2)}, &\textrm{for $ \Omega= 2k\pi k_BT $}.
\end{array} \right.\nonumber\\
\bar{\kappa}(\omega_{mn})&&=\int_0^{\infty}d\tau \mu (\tau)\cos(\omega_{mn}\tau)=\frac{\eta\pi\Omega^2}{2(\Omega^2+\omega_{mn}^2)}, \nonumber \\
\bar{\gamma}(\omega_{mn})&&=\int_0^{\infty}d\tau\mu \sin(\omega_{mn}\tau) \nonumber\\
&&= \left\{\begin{array}{ll}
      \frac{\pi}{2}J(\omega_{mn}),  &  \textrm{for $\omega_{mn} \neq 0 $}, \\
      0,  &  \textrm{for $\omega_{mn} = 0 $}. \\
  \end{array} \right. \label{dfrk}
\end{eqnarray}
Here, $\bar{\omega}_j=2j\pi k_B T, (j=1,2,...,N)$. It is easy to be verified that
\begin{eqnarray}
\left\{\begin{array}{ll}
      \Gamma_{\alpha\beta}^{(1)}(\omega_{mn})=Re\left[ \Gamma_{\alpha\beta}^{(2)}(\omega_{mn})\right] ,  &  \textrm{for $\omega_{mn} \neq 0 $}, \\
      \Gamma_{\alpha\beta}^{(1)}(\omega_{mn})=\mathop{lim}\limits_{\Omega/k_BT\to 0}Re\left[ \Gamma_{\alpha\beta}^{(2)}(\omega_{mn})\right], & \textrm{for $\omega_{mn} = 0 $}.
\end{array} \right. \nonumber\\
\end{eqnarray}
It is shown that $\Gamma^{(1)}_{\alpha\beta}(\omega_{mn}) $ is the real part of the $\Gamma^{(2)}_{\alpha\beta}(\omega_{mn})$ in high temperature approximation. In the following we shall investigate the reduced dynamics of two models by using the non-secular Lindblad and Redfield master equations and $\Gamma^{(2)}_{\alpha\beta}(\omega_{mn})$. The dynamics obtained from the secular Lindblad and Redfield master equation and $\Gamma^{(1)}_{\alpha\beta}(\omega_{mn})$ are also investigated. The two kinds of results will be compared each other.

\section{two example}
In this section we shall investigate the dynamics of an open three-level quantum system model and PE545 complex model by using Lindblad and Redfield master equations Eqs.(\ref{Lindblad}), (\ref{Lindbladse}) and (\ref{Redfield}) with $\Gamma^{(1)}_{\alpha\beta}(\omega_{mn}) $ and $\Gamma^{(2)}_{\alpha\beta}(\omega_{mn})$.

\subsection{An open three-level quantum system model}
At first, we investigate an open three-level quantum system model. The system's Hamiltonian is set as
\begin{eqnarray}
  H_s=\left(\begin{array}{ccc}
      E_1     & V_{12}  & 0      \\
      V_{21}  & E_2     & V_{23} \\
      0       & V_{32}  & E_3
  \end{array}\right),
\end{eqnarray}
where we set $E_1=0, E_2=-2.67$cm$^{-1}, E_3=-3.67$cm$^{-1} $, and $V_{12}=V_{21}=V_{23}=V_{32}=0.67$cm$^{-1} $, and the environment is described with the Drude spectral density function as Eq.(\ref{J}). Supposing that $\eta=0.125, \Omega=100.0$cm$^{-1} $, and the environmental temperature is 300K. The initial state is set as $\rho(0)=\vert 1\rangle\langle 1\vert $.  Numerical analysis shows that when $N>=100$, $\bar{D}$, and $\bar{f}$ are convergence, so we set $N=100$ in Eq.(\ref{dfrk}), here. We solve the reduced dynamics of the open three-level system by using the non-secular Lindblad equation with $\Gamma^{(2)}_{\alpha\beta}(\omega_{mn})$, see Fig.1(a), and $\Gamma^{(1)}_{\alpha\beta}(\omega_{mn})$, see Fig.1(c); non-secular Redfield equation with $\Gamma^{(2)}_{\alpha\beta}(\omega_{mn})$, see Fig.1(b), and  $\Gamma^{(1)}_{\alpha\beta}(\omega_{mn})$, see Fig.1(d); secular Lindblad equation with $\Gamma^{(2)}_{\alpha\beta}(\omega_{mn})$, see Fig.1(e), and $\Gamma^{(1)}_{\alpha\beta}(\omega_{mn})$, see Fig.1(g); secular Redfield equation with $\Gamma^{(2)}_{\alpha\beta}(\omega_{mn})$, see Fig.1(f), and $\Gamma^{(1)}_{\alpha\beta}(\omega_{mn})$, see Fig.1(h).
\begin{figure}[ht]
\includegraphics[width=0.5\textwidth]{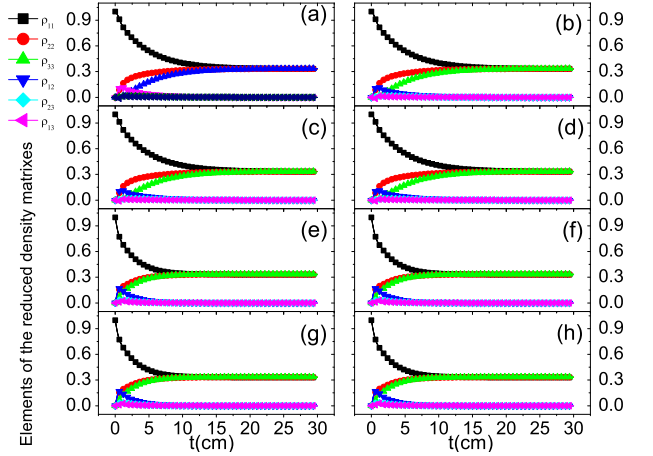}
\vspace{0.0em}
\caption{\label{Fig.1}The evolution of reduced density matrix for a tree-level quantum system obtained  by using the non-secular Lindblad equation with $\Gamma^{(2)}_{\alpha\beta}(\omega_{mn})$ (a), and $\Gamma^{(1)}_{\alpha\beta}(\omega_{mn})$ (c); non-secular Redfield equation with $\Gamma^{(2)}_{\alpha\beta}(\omega_{mn})$ (b), and  $\Gamma^{(1)}_{\alpha\beta}(\omega_{mn})$ (d); secular Lindblad equation with $\Gamma^{(2)}_{\alpha\beta}(\omega_{mn})$ (e), and  $\Gamma^{(1)}_{\alpha\beta}(\omega_{mn})$ (g); secular Redfield equation with $\Gamma^{(2)}_{\alpha\beta}(\omega_{mn})$ (f), and  $\Gamma^{(1)}_{\alpha\beta}(\omega_{mn})$ (h). Here,  the environment is described with the Drude spectral density function, with $\eta=0.125, \Omega=100.0$cm$^{-1} $, $T= 300$K. The initial state is set as $\rho(0)=\vert 1\rangle\langle 1\vert $. In Eq.(\ref{dfrk}) $N=100$.}
\end{figure}
It is shown that for the three-level quantum system, the distortion of quantum disspative dynamics due to use of the secular approximation and  $\Gamma^{(1)}_{\alpha\beta}(\omega_{mn})$ is not serious, and a little distortion does not result from the secular approximation but from the usage of inexact $\Gamma^{(1)}_{\alpha\beta}(\omega_{mn})$. The results obtained from the non-secular Lindblad and Redfield equations with $\Gamma^{(2)}_{\alpha\beta}(\omega_{mn})$ are exactly agreement with the ones obtained from the Born-Markov equation \cite{PRA99}.

\subsection{PE545 model}
Secondly, we investigate the reduced dynamics of the phycoerythrin 545 (PE545) from Rhodomonas CS24 \cite{nature463,JPCB118}. The model Hamiltonian of the PE545 is
\begin{widetext}
\begin{eqnarray}
  H_s=\left(\begin{array}{cccccccc}
      18008. & -4.1  & -31,9 & 2.8  & 2.1   & -37.1  & -10.5   & 45.9 \\
      -4.1   & 17973 & -2.9  & 30.9 & -35.9 & 2.5    & -45.5   & 11.0 \\
      -31.9  & -2.9  & 18711 & -5.6 & -19.6 & -16.1  & 6.7     & 6.8  \\
      2.8    & 30.9  & -5.6  & 18960& 11.5  & 25.5   & 5.1     & 7.4  \\
      2.1    & -35.4 & -19.6 & 11.5 & 18532 & 101.5  & 36.3    & 16.0 \\
      -37.1  & 2.5   & -16.1 & 25.5 & 101.5 & 19574  & 17.6    & -38.6\\
      -10.5  & -45.5 & 6.7   & 5.1  & 36.3  & 17.6   & 18040   & 2.6  \\
      45.9   & 11.0  & 6.8   & 7.4  & 16.0  & -38.6  & 2.6     & 19050
  \end{array}\right).
\end{eqnarray}
\end{widetext}
Here, the environment is also described with the Drude spectral density function, and we set $\eta=12.5,\Omega=1000$ cm$^{-1} $, and suppose the environmental temperature is 300K. The initial state is set as $\rho(0)=\vert 1\rangle\langle 1\vert$.  When $N>=10000$, $\bar{D}$, and $\bar{f}$ are convergence, so we set $N=10000$ in Eq.(\ref{dfrk}), here. We obtain the evolution of the elements of reduced density matrix for the PE545 model by using the non-secular Lindblad equation with $\Gamma^{(2)}_{\alpha\beta}(\omega_{mn}) $, see Fig.2(a), and $\Gamma^{(1)}_{\alpha\beta}(\omega_{mn})$, see Fig.2(c); non-secular Redfield equation with $\Gamma^{(2)}_{\alpha\beta}(\omega_{mn})$, see Fig.2(b), and  $\Gamma^{(1)}_{\alpha\beta}(\omega_{mn})$, see Fig.2(d); secular Lindblad equation with $\Gamma^{(2)}_{\alpha\beta}(\omega_{mn})$, see Fig.2(e), and  $\Gamma^{(1)}_{\alpha\beta}(\omega_{mn})$, see Fig.2(g); secular Redfield equation with $\Gamma^{(2)}_{\alpha\beta}(\omega_{mn})$, see Fig.2(f), and  $\Gamma^{(1)}_{\alpha\beta}(\omega_{mn})$, see Fig.2(h). From the evolution of the elements of the density matrix for the PE545, we see that both the secular approximation and the simplified $\Gamma_{\alpha\beta}^{(1)}(\omega_{mn})$ affect the dynamics of the complex open quantum system. The secular approximation makes the time scale of system evolution about ten times longer, and the simplified $\Gamma^{(1)}_{\alpha\beta}(\omega_{mn}) $ compresses the time scale of system evolution by about 100 times which are seriously distorts the actual dynamics of the complex open quantum system. It can be imagined that for systems with different degrees of complexity, the time scale distortion of dynamics due to use of the secular approximation with simplified spectral correlation tensor of bath $\Gamma^{(1)}_{\alpha\beta}(\omega_{mn})$ will also be different. For more complex systems, this distortion will also be more serious.
\begin{figure}[ht]
\includegraphics[width=0.5\textwidth]{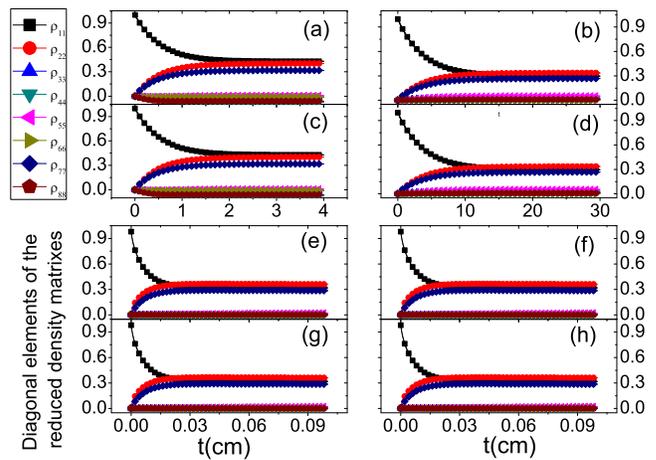}
\vspace{3.0em}
\caption{\label{Fig.2}The evolution of reduced density matrix for  PE545 model obtained  by using the non-secular Lindblad and Redfield equations, where the meanings of (a)-(h) are similar to them in Fig.1.  And,  the environment is described with the Drude spectral density function, with $\eta=12.5, \Omega=1000.0$ cm$^{-1} $, $T= 300$K. The initial state is set as $\rho(0)=\vert 1\rangle\langle 1\vert $. In Eq.(\ref{dfrk}), $N=10000$.}
\end{figure}

\section{Discussion and conclusions}
In this paper we re-derive the computable forms of Lindblad-like and Radfield-like master equations, and reevaluate the corresponding coefficients in the equations. And by using the secular and non-secular, Lindblad and Redfield master equations with different representations of the spectral correlation tensor of bath $\Gamma^{(1)}_{\alpha\beta}(\omega_{mn})$, and $\Gamma^{(2)}_{\alpha\beta}(\omega_{mn})$, we investigated two quantum system models. One is an open three-level quantum system model, and the other is the PE545 model. It is shown that the secular approximation and simplified spectral correlation tensor bath $\Gamma^{(1)}_{\alpha\beta}(\omega_{mn})$ seriously affect the accuracy of the results obtained from Lindblad-like and Redfield-like master equations, especially for some complex open quantum systems.

\section{acknowledgements}

This project was sponsored by the National Natural Science Foundation of China (Grant No. 21773131), and the K.C. Wong Magna Foundation in Ningbo University.

\end{document}